\documentclass[aps,twocolumn,showkeys,showpacs,preprintnumbers,prd,superscriptaddress,nofootinbib,10pt]{revtex4-1}
\usepackage{graphicx,epsf,bm,amsmath,amsfonts,amssymb,epstopdf,natbib,hyperref,color,verbatim,multirow,bm}
\hypersetup{colorlinks=true,urlcolor=blue,citecolor=blue,linkcolor=blue,menucolor=blue,anchorcolor=blue,filecolor=blue}
\usepackage{appendix}
\date{\today}

\begin{document}

\title{Primordial regular black holes as all the dark matter. \\ III. Covariant canonical quantum gravity models}

\author{Marco Calz\`{a}}
\email{marco.calza@unitn.it}
\affiliation{Department of Physics, University of Trento, Via Sommarive 14, 38123 Povo (TN), Italy}
\affiliation{Trento Institute for Fundamental Physics and Applications (TIFPA)-INFN, Via Sommarive 14, 38123 Povo (TN), Italy}

\author{Davide Pedrotti}
\email{davide.pedrotti-1@unitn.it}
\affiliation{Department of Physics, University of Trento, Via Sommarive 14, 38123 Povo (TN), Italy}
\affiliation{Trento Institute for Fundamental Physics and Applications (TIFPA)-INFN, Via Sommarive 14, 38123 Povo (TN), Italy}

\author{Guan-Wen Yuan}
\email{guanwen.yuan@unitn.it}
\affiliation{Department of Physics, University of Trento, Via Sommarive 14, 38123 Povo (TN), Italy}
\affiliation{Trento Institute for Fundamental Physics and Applications (TIFPA)-INFN, Via Sommarive 14, 38123 Povo (TN), Italy}

\author{Sunny Vagnozzi}
\email{sunny.vagnozzi@unitn.it}
\affiliation{Department of Physics, University of Trento, Via Sommarive 14, 38123 Povo (TN), Italy}
\affiliation{Trento Institute for Fundamental Physics and Applications (TIFPA)-INFN, Via Sommarive 14, 38123 Povo (TN), Italy}

\begin{abstract}
\noindent In earlier companion papers, we showed that non-singular primordial black holes (PBHs) could account for all the dark matter (DM) over a significantly wider mass range compared to Schwarzschild PBHs. Those studies, mostly based on phenomenological metrics, are now extended by considering the quantum-corrected space-time recently proposed by Zhang, Lewandowski, Ma \& Yang (ZLMY), derived from an effective canonical (loop) quantum gravity approach explicitly enforcing general covariance. Unlike the BHs considered earlier, ZLMY BHs are free from Cauchy horizons, and are hotter than their Schwarzschild counterparts. We show that this higher temperature boosts the evaporation spectra of ZLMY PBHs, tightening limits on their abundance relative to Schwarzschild PBHs and shrinking the asteroid mass window where they can constitute all the DM, a result which reverses the earlier trend, but rests on firmer theoretical ground. While stressing the potential key role of quantum gravity effects in addressing the singularity and DM problems, our study shows that working within a consistent theoretical framework can strongly affect observational predictions.
\end{abstract}

\maketitle

\section{Introduction}
\label{sec:introduction}

Three among the most important open questions in theoretical physics are arguably the nature of the dark matter (DM)~\cite{Arbey:2021gdg,Cirelli:2024ssz}, the singularity problem which plagues General Relativity (GR)~\cite{Penrose:1964wq,Hawking:1970zqf}, and the quest for a framework of quantum gravity (QG)~\cite{Addazi:2021xuf}. Quite remarkably, black holes (BHs) not only offer a concrete setting to test possible solutions to the singularity problem (in no small part thanks to a wide array of observations providing novel insight into fundamental physics in the strong-field regime~\cite{Creminelli:2017sry,Sakstein:2017xjx,Ezquiaga:2017ekz,Boran:2017rdn,Baker:2017hug,Amendola:2017orw,Visinelli:2017bny,Crisostomi:2017lbg,Dima:2017pwp,Cai:2018rzd,Casalino:2018tcd,Barack:2018yly,LIGOScientific:2018dkp,Casalino:2018wnc,Held:2019xde,Bambi:2019tjh,Vagnozzi:2019apd,Zhu:2019ura,Cunha:2019ikd,Banerjee:2019nnj,Banerjee:2019xds,Allahyari:2019jqz,Vagnozzi:2020quf,Khodadi:2020jij,Kumar:2020yem,Khodadi:2020gns,Pantig:2021zqe,Khodadi:2021gbc,Roy:2021uye,Uniyal:2022vdu,Pantig:2022ely,Ghosh:2022kit,Khodadi:2022pqh,KumarWalia:2022aop,Shaikh:2022ivr,Odintsov:2022umu,Oikonomou:2022tjm,Pantig:2023yer,Gonzalez:2023rsd,Sahoo:2023czj,Nozari:2023flq,Uniyal:2023ahv,Filho:2023ycx,Raza:2023vkn,Hoshimov:2023tlz,Chakhchi:2024tzo,Liu:2024lbi,Liu:2024lve,Khodadi:2024ubi,Nojiri:2024txy}), but may also be the key towards addressing the QG and DM puzzles. On the one hand, in fact, reconciling quantum mechanics with the existence of BHs, specifically their apparent non-unitary evolution~\cite{Hawking:1975vcx}, is a critical problem any QG framework ultimately has to address. On the other hand, primordial BHs (PBHs) formed in the early Universe from the collapse of large density perturbations re-entering the horizon, are among the most promising DM candidates and have gained widespread interest over the past decade~\cite{Chapline:1975ojl,Meszaros:1975ef,Khlopov:1980mg,Khlopov:1985fch,Ivanov:1994pa,Choudhury:2013woa,Belotsky:2014kca,Bird:2016dcv,Clesse:2016vqa,Poulin:2017bwe,Raccanelli:2017xee,LuisBernal:2017fmf,Clesse:2017bsw,Pi:2017gih,Kohri:2018qtx,Cai:2018dig,Liu:2018ess,Liu:2019rnx,Murgia:2019duy,Carr:2019kxo,Liu:2020cds,Hertzberg:2020hsz,Serpico:2020ehh,DeLuca:2020bjf,DeLuca:2020fpg,DeLuca:2020qqa,Carr:2020erq,Bhagwat:2020bzh,DeLuca:2020sae,Wong:2020yig,Carr:2020mqm,Domenech:2020ssp,DeLuca:2021wjr,Arbey:2021ysg,Franciolini:2021tla,DeLuca:2021hde,Cheek:2021odj,Cheek:2021cfe,Mittal:2021egv,Heydari:2021gea,Dvali:2021byy,Heydari:2021qsr,DeLuca:2021pls,Liu:2021jnw,Saha:2021pqf,Pi:2021dft,Bhaumik:2022pil,Anchordoqui:2022txe,Cai:2022erk,Oguri:2022fir,Franciolini:2022tfm,Mazde:2022sdx,Cai:2022kbp,Anchordoqui:2022tgp,Liu:2022iuf,Ferrante:2022mui,Fu:2022ypp,Choudhury:2023vuj,Papanikolaou:2023crz,Choudhury:2023jlt,Choudhury:2023rks,Musco:2023dak,Yuan:2023bvh,Choudhury:2023hvf,Ghoshal:2023sfa,Cai:2023uhc,Ferrante:2023bgz,Choudhury:2023kdb,Huang:2023chx,Choudhury:2023hfm,Bhattacharya:2023ysp,Heydari:2023xts,Heydari:2023rmq,Choudhury:2023fwk,Choudhury:2023fjs,Ghoshal:2023pcx,Hai-LongHuang:2023atg,Huang:2023mwy,Anchordoqui:2024akj,Choudhury:2024one,Ianniccari:2024bkh,Thoss:2024hsr,Papanikolaou:2024kjb,Choudhury:2024ybk,Choudhury:2024jlz,Anchordoqui:2024dxu,Wang:2024vfv,Papanikolaou:2024fzf,Yin:2024xov,Barroso:2024cgg,Andres-Carcasona:2024wqk,Choudhury:2024dei,Heydari:2024bxj,Dvali:2024hsb,Boccia:2024nly,Iovino:2024tyg,Huang:2024aog,Choudhury:2024dzw,Anchordoqui:2024jkn,Yang:2024vij,Saha:2024ies,He:2024luf,Anchordoqui:2024tdj,Chen:2024pge,Dai:2024guo,Choudhury:2024kjj,Hai-LongHuang:2024vvz,Zantedeschi:2024ram,Chianese:2024rsn,Barker:2024mpz,Borah:2024bcr,Hai-LongHuang:2024gtx,Ahmed:2024tlw,Athron:2024fcj,Yang:2024pfb,Allegrini:2024ooy,Zhao:2024jad,Wang:2024nmd,Yogesh:2025hll,Calabrese:2025sfh,Wang:2025hbw,Crescimbeni:2025ywm,Ghoshal:2025dmi,Hai-LongHuang:2025vfs,Liu:2025vpz,Khan:2025kag,Montefalcone:2025akm,Dvali:2025ktz,Yang:2025uvf,Jia:2025vqn,Anchordoqui:2025xug,Jiang:2025jxt,Dondarini:2025ktz,Anchordoqui:2025opy,Zhao:2025ddy,DeLuca:2025uov,Kumar:2025jfi,Papanikolaou:2025ddc} (see e.g.\ Refs.~\cite{Green:2020jor,Bird:2022wvk,Carr:2023tpt,Arbey:2024ujg,Choudhury:2024aji} for recent reviews). In the rest of this work, PBHs and observational tests of their possible existence will serve as a bridge between the DM, singularity, and QG problems.

Nearly all earlier studies on PBHs as DM candidates worked under the assumption of these objects being Schwarzschild or Kerr PBHs. Under such an assumption, it has been found that PBHs can account for the entire DM component of the Universe only within the so-called ``\textit{asteroid mass window}'', for PBH masses $10^{17}\,{\text{g}} \lesssim M_{\text{pbh}} \lesssim 10^{23}\,{\text{g}}$: lighter PBHs are tightly constrained by the non-observation of products of their Hawking evaporation, whereas heavier ones are strongly constrained by the absence of microlensing signatures expected from their presence in the galactic halo~\cite{Bai:2018bej,Smyth:2019whb,Coogan:2020tuf,Ray:2021mxu,Auffinger:2022dic,Ghosh:2022okj,Miller:2021knj,Branco:2023frw,Bertrand:2023zkl,Tran:2023jci,Gorton:2024cdm,Dent:2024yje,Tamta:2024pow,Tinyakov:2024mcy,Loeb:2024tcc,Li:2025mqx}. All observational constraints on PBHs, including those determining the extension of the asteroid mass window, are contingent on the assumption of Schwarzschild and/or Kerr PBHs. While on the phenomenological side this is a reasonable assumption, on the theoretical side the situation is quite different, given that both space-times feature well-known curvature singularities. These considerations motivated some of us, in two earlier companion papers~\cite{Calza:2024fzo,Calza:2024xdh}, to entertain the possibility that DM is composed of primordial \textit{regular} (non-singular) BHs (PRBHs):~\footnote{Specifically, in the first work~\cite{Calza:2024fzo} we considered three \textit{tr} (time-radial)-symmetric regular space-times (the Bardeen~\cite{Bardeen:1968ghw}, Hayward~\cite{Hayward:2005gi}, and Culetu-Ghosh-Simpson-Visser ones~\cite{Culetu:2013fsa,Culetu:2014lca,Ghosh:2014pba,Simpson:2019mud}), whereas the second~\cite{Calza:2024xdh} studied three non-\textit{tr}-symmetric metrics (the Simpson-Visser~\cite{Simpson:2018tsi}, Peltola-Kunstatter~\cite{Peltola:2008pa,Peltola:2009jm}, and D'Ambrosio-Rovelli ones~\cite{Bianchi:2018mml,DAmbrosio:2018wgv}). See e.g.\ Refs.~\cite{Easson:2002tg,Dymnikova:2015yma,Pacheco:2018mvs,Arbey:2021mbl,Arbey:2022mcd,deFreitasPacheco:2023hpb,Banerjee:2024sao,Davies:2024ysj,Dialektopoulos:2025mfz,Trivedi:2025vry,Carr:2025auw,Asmanoglu:2025agc,Jusufi:2025qgd,Khodadi:2025icd,Trivedi:2025agk,Loc:2025mzc} for other works considering the possibility that primordial non-Schwarzschild/Kerr objects, and/or remnants thereof, may play a role in the DM problem.} in short, we found that evaporation constraints on all the six PRBH space-times studied are significantly relaxed compared to their Schwarzschild counterparts (and can actually be completely evaded under certain extremality assumptions), resulting in a much larger window in mass where PRBHs can account for all of the DM compared to the Schwarzschild PBHs case.

Why, then, are we now considering a third installment in the ``primordial regular black hole'' line of work? Two key considerations motivate our study, one being primarily phenomenological, whereas the other is more theoretical in nature.  To begin with, in all six space-times analyzed in our earlier works, the BH temperature (at a given mass) was consistently lower than the Schwarzschild value, vanishing in the extremal limit (see e.g.\ Fig.~1 in Refs.~\cite{Calza:2024fzo,Calza:2024xdh}): this behaviour is directly responsible for the weaker evaporation constraints on these PRBHs, given the exponential dependence of the Hawking radiation spectrum on the BH temperature. In Ref.~\cite{Calza:2024fzo} (see footnote~2 thereof), we argued that this decreasing temperature behaviour could be quite generic, but left open the possibility that some regular BHs may exhibit higher temperatures. Additionally, the space-times considered previously were either phenomenological~\cite{Bardeen:1968ghw,Hayward:2005gi,Culetu:2013fsa,Culetu:2014lca,Ghosh:2014pba,Simpson:2019mud,Simpson:2018tsi}, or inspired by Loop Quantum Gravity (LQG)~\cite{Peltola:2008pa,Peltola:2009jm,Bianchi:2018mml,DAmbrosio:2018wgv}. However, in this second case, the metrics considered were typically constructed using simplified, at times heuristic assumptions. Perhaps more importantly, these constructions typically do not explicitly address the issue of general covariance: this is a crucial consistency requirement for any effective canonical QG theory, but is usually not manifest in these theories (as we elaborate in detail later on), and has been the subject of a long debate~\cite{Bojowald:2008gz,Tibrewala:2013kba,Bojowald:2015zha,Wu:2018mhg,Bojowald:2020xlw,Bojowald:2020unm,Gambini:2022dec,Bojowald:2022zog,Han:2022rsx,Giesel:2023hys}. This raises the important question of whether LQG-inspired space-times constructed with a clearer link to the full framework, and at the same time ensuring general covariance, might lead to different physical predictions.

In this work, we focus on the quantum-corrected regular BH solution first presented in Ref.~\cite{Zhang:2024ney}, which we refer to as ``Zhang-Lewandowski-Ma-Yang'' (ZLMY) BH. This solution, which has gained tremendous interest recently~\cite{Feng:2024sdo,Konoplya:2024lch,Liu:2024soc,Liu:2024wal,Malik:2024nhy,Heidari:2024bkm,Wang:2024iwt,Skvortsova:2024msa,Ban:2024qsa,Du:2024ujg,Lin:2024beb,Shu:2024tut,Liu:2024pui,Liu:2024iec,Paul:2025wen,Konoplya:2025hgp,Chen:2025ifv,Xamidov:2025oqx,Yang:2025ufs,Ai:2025myf,Wang:2025alf,Lutfuoglu:2025hwh,Chen:2025aqh,Sahlmann:2025fde,Zhang:2025ccx}, explicitly addresses the two key issues identified earlier: its temperature increases relative to the Schwarzschild case, and it emerges from an approach which explicitly enforces general covariance in the LQG framework (therefore ensuring the soundness of the effective canonical theory)~\cite{Zhang:2024khj}. Last but not least, unlike most regular BH solutions in the literature, the ZLMY BH is free from Cauchy horizons, and is therefore safe against perturbative instabilities. Extending our earlier studies~\cite{Calza:2024fzo,Calza:2024xdh} to consider primordial ZLMY BHs as DM candidates, we find that the evaporation constraints on their abundance are tighter because of their increased temperature, naturally leading to a smaller window where these BHs can account for all the DM. Overall, our study naturally ties together the DM, singularity, and QG problems in a way which is theoretically more robust and well-grounded compared to earlier works~\cite{Calza:2024fzo,Calza:2024xdh}, while leading to rather distinct observational predictions.

The rest of this paper is then organized as follows. In Sec.~\ref{sec:zlmy}, after briefly discussing the issue of general covariance in effective canonical QG theories, we introduce the ZLMY BH and calculate its temperature. The methodology adopted to calculate the resulting Hawking evaporation spectra and compare these against astrophysical observations is presented in Sec.~\ref{sec:methodology}. Our main results, which consist in limits on the fraction of DM in the form of ZLMY PBHs, are obtained and critically discussed in Sec.~\ref{sec:results}. Finally, in Sec.~\ref{sec:conclusions} we draw concluding remarks. Explicit expressions for the curvature invariants of the ZLMY BH are instead provided in Appendix~\ref{sec:invariants}. Unless otherwise specified, we work in units where $G=c=\hbar=1$. Finally, we recall once more that this is the third part of a series of pilot studies on the possibility of primordial regular BHs making up all the DM. Readers may find it beneficial to consult our two previous companion papers in this series~\cite{Calza:2024fzo,Calza:2024xdh}, though this is by no means necessary, as we have taken care to ensure that the present work is self-contained.

\section{Zhang-Lewandowski-Ma-Yang regular black hole}
\label{sec:zlmy}

The Penrose-Hawking singularity theorems have established the unavoidable appearance of singularities in GR~\cite{Penrose:1964wq,Hawking:1970zqf} (see Refs.~\cite{Ong:2020xwv,deHaro:2023lbq,Trivedi:2023zlf} for recent reviews). These singularities are clearly undesirable, since they signal a breakdown of predictability (see, however, Refs.~\cite{Sachs:2021mcu,Ashtekar:2021dab,Ashtekar:2022oyq} for a different view on this problem): for this reason, significant effort has been devoted to the study of so-called regular BHs, i.e.\ space-times which are free of singularities~\cite{Borde:1996df,AyonBeato:1998ub,AyonBeato:1999rg,Bronnikov:2005gm,Berej:2006cc,Bronnikov:2012ch,Rinaldi:2012vy,Stuchlik:2014qja,Schee:2015nua,Johannsen:2015pca,Myrzakulov:2015kda,Fan:2016hvf,Sebastiani:2016ras,Toshmatov:2017zpr,Chinaglia:2017uqd,Frolov:2017dwy,Bertipagani:2020awe,Nashed:2021pah,Simpson:2021dyo,Franzin:2022iai,Chataignier:2022yic,Ghosh:2022gka,Khodadi:2022dyi,Farrah:2023opk,Fontana:2023zqz,Boshkayev:2023rhr,Luongo:2023jyz,Luongo:2023aib,Cadoni:2023lum,Giambo:2023zmy,Cadoni:2023lqe,Luongo:2023xaw,Sajadi:2023ybm,Javed:2024wbc,Ditta:2024jrv,Al-Badawi:2024lvc,Ovgun:2024zmt,Corona:2024gth,Bueno:2024dgm,Konoplya:2024hfg,Pedrotti:2024znu,Bronnikov:2024izh,Kurmanov:2024hpn,Capozziello:2024ucm,Bolokhov:2024sdy,Agrawal:2024wwt,Belfiglio:2024wel,Stashko:2024wuq,Faraoni:2024ghi,Khodadi:2024efq,Calza:2024qxn,Estrada:2024uuu,KumarWalia:2024yxn,Li:2024ctu,Estrada:2024moz,Benavides-Gallego:2024hck,Frolov:2024hhe,Balart:2024rtj,Bueno:2024zsx,Bueno:2024eig,Vertogradov:2025snh,Sajadi:2025prp,Xiong:2025hjn,Estrada:2025aeg,Casadio:2025pun,Harada:2025cwd,Bueno:2025dqk,Fauzi:2025ldu,Kala:2025xnb,Capozziello:2025ycu,Pedrotti:2025idg,Urmanov:2025nou,Pinto:2025loq,Calza:2025mrt,Bueno:2025gjg,Neves:2025uoi,Capozziello:2025wwl,Eichhorn:2025pgy,Bonanno:2025dry,Bueno:2025zaj,Calza:2025yfm}, see Refs.~\cite{Sebastiani:2022wbz,Lan:2023cvz,Carballo-Rubio:2025fnc} for reviews.~\footnote{It is worth clarifying that finiteness of curvature invariants is not in itself a guarantee of geodesic completeness. This is a problem which affects a number of well-known regular BH solutions~\cite{Zhou:2022yio}. In addition, an issue currently under debate concerns whether several regular BH solutions are unstable against perturbations, due to the so-called mass inflation instability~\cite{Carballo-Rubio:2021bpr,Carballo-Rubio:2022kad,Bonanno:2022jjp,Bonanno:2022rvo,Carballo-Rubio:2022twq}.} In this work, following up on our earlier companion papers~\cite{Calza:2024fzo,Calza:2024xdh}, we will consider the possibility of primordial regular BHs making up the DM. The generic space-time we consider is characterized by a line element taking the following form:
\begin{equation}
ds^2 = -f(r,\ell)dt^2+g(r,\ell)^{-1}dr^2+r^2d\Omega^2\,,
\label{eq:ds2}
\end{equation}
where $\ell$ is a regularizing parameter, such that the Schwarzschild metric is typically recovered in the $\ell \to 0$ limit. We require asymptotic flatness, which imposes the following constraints on the functions $f(r)$ and $g(r)$:
\begin{equation}
f(r) \xrightarrow{r \to \infty} 1\,, \quad g(r) \xrightarrow{r \to \infty} 1\,.
\label{eq:asflat}
\end{equation}
In the case we will study, i.e.\ the ZLMY BH, $f(r) \neq g(r)$, so the metric is a non-\textit{tr}-symmetric one.

It is widely believed that the singularity problem in GR should ultimately be cured by quantum gravity effects, although only a few first-principles studies concretely support this belief~\cite{Dymnikova:1992ux,Dymnikova:2004qg,Ashtekar:2005cj,Bebronne:2009mz,Modesto:2010uh,Spallucci:2011rn,Perez:2014xca,Colleaux:2017ibe,Nicolini:2019irw,Bosma:2019aiu,Jusufi:2022cfw,Olmo:2022cui,Jusufi:2022rbt,Ashtekar:2023cod,Nicolini:2023hub}. Progress in this direction is not easily achieved, in part because the ultimate QG framework remains unknown, making it particularly challenging to incorporate QG effects. One possibility for investigating possible QG effects is to treat the unknown QG theory as an effective field theory~\cite{Donoghue:2012zc}. A specific framework within this approach is effective canonical QG, where one attempts to quantize the canonical formulation of the effective QG theory, resulting in a (semiclassical) model of gravity governed by an effective Hamiltonian constraint. While promising, this approach raises an important question: under what conditions does a given $3+1$ model in the Hamiltonian formulation maintain 4-dimensional diffeomorphism covariance, i.e.\ describes a generally covariant theory? While in a Lagrangian formulation this issue is straightforwardly addressed by ensuring that the action is generally covariant, the $3+1$ space-time decomposition in the Hamiltonian formulation makes the task challenging. This \textit{covariance issue} is an important problem for any effective canonical QG theory~\cite{Bojowald:2008gz,Tibrewala:2013kba,Bojowald:2015zha,Wu:2018mhg,Bojowald:2020xlw,Bojowald:2020unm,Gambini:2022dec,Bojowald:2022zog,Han:2022rsx,Giesel:2023hys}: while it has been investigated in various works, the pioneering study of Zhang, Lewandowski, Ma, and Yang~\cite{Zhang:2024khj} took a significant step forward, addressing the problem without resorting to a specific gauge-fixing procedure, and focusing on solutions related to loop quantum BH models, i.e.\ symmetry-reduced sectors of loop quantum gravity (LQG). Briefly stated, Ref.~\cite{Zhang:2024khj} shows that for an effective canonical QG theory to be generally covariant, the effective Hamiltonian has to take a specific form, controlled by an ``effective mass'' $M_{\text{eff}}$, itself required to be a solution to two ``covariance equations'' -- see Eqs.~(7a,7b) of Ref.~\cite{Zhang:2024khj}. Solutions to the covariance equations lead to distinct effective masses $M_{\text{eff}}$, and hence different effective Hamiltonians $H_{\text{eff}}$, each defining an effective QG theory which is \textit{guaranteed} to be generally covariant.

The above procedure was then applied to LQG-related models in Refs.~\cite{Zhang:2024khj,Zhang:2024ney}. We recall that LQG is a background-independent, non-perturbative approach to quantizing GR, characterized by the polymerization procedure which replaces connections with their holonomies, resulting in a discrete quantum geometry (for a review, see e.g.\ Ref.~\cite{Ashtekar:2021kfp}). In Ref.~\cite{Zhang:2024khj}, it is argued that within the approach at hand the holonomies reduce to combinations of trigonometric functions of the extrinsic curvature. A generic polymerization therefore consists in replacing the extrinsic curvature with a trigonometric function thereof, with this function controlled by a new quantum (regularizing) parameter. It is worth emphasizing how this procedure is radically different from the usual one, where the polymerization procedure is applied directly on the classical Hamiltonian constraint, resulting in models whose general covariance is not guaranteed.

As a direct application of the above methodology, Refs.~\cite{Zhang:2024khj,Zhang:2024ney} derived three solutions to the covariance equations, leading to three distinct new effective Hamiltonian constraints. From these, three non-rotating (static, spherically symmetric) space-times have been obtained, and found to be closely related to earlier loop quantum BH models, confirming the close connection between the corresponding covariant effective canonical QG models and LQG: these solutions have gained significant interest recently, see e.g.\ Refs.~\cite{Feng:2024sdo,Konoplya:2024lch,Liu:2024soc,Liu:2024wal,Malik:2024nhy,Heidari:2024bkm,Wang:2024iwt,Skvortsova:2024msa,Ban:2024qsa,Du:2024ujg,Lin:2024beb,Shu:2024tut,Liu:2024pui,Liu:2024iec,Paul:2025wen,Konoplya:2025hgp,Chen:2025ifv,Xamidov:2025oqx,Yang:2025ufs,Ai:2025myf,Wang:2025alf,Lutfuoglu:2025hwh,Chen:2025aqh,Sahlmann:2025fde,Zhang:2025ccx} for studies in this direction. Of particular interest to us is the third solution to the covariance equations, derived in Ref.~\cite{Zhang:2024ney}. This is actually a family of solutions controlled by an integer parameter $n$, and the line element of the associated static spherically symmetric space-time, which we refer to as $ds^2_{(3)}$, is given by the following:
\begin{equation}
ds^2_{(3)} = -\bar{f}_3^{(n)}ds^2+\bar{\mu}_3^{-1} \left ( \bar{f}_3^{(n)} \right ) ^{-1}dr^2+r^2d\Omega^2\,,
\label{eq:ds23}
\end{equation}
with $\bar{f}_3^{(n)}(r)$ is labeled by an integer index $n \in \mathbb{Z}$ and given by the following expression, where for clarity we have momentarily restored Newton's constant $G$:
\begin{equation}
\bar{f}_3^{(n)}(r) = 1-(-1)^n\frac{r^2}{\xi^2}\arcsin \left ( \frac{2GM\xi^2}{r^3} \right ) -\frac{n\pi r^2}{\xi^2}\,,
\label{eq:f3}
\end{equation}
whereas $\bar{\mu}_3(r)$ is given by the following (again temporarily restoring $G$):
\begin{equation}
\bar{\mu}_3(r) = 1-\frac{4G^2\xi^4M^2}{r^6}\,.
\label{eq:mu3}
\end{equation}
In both Eqs.~(\ref{eq:f3},\ref{eq:mu3}), $\xi \propto \sqrt{\hbar}$ is a quantum parameter proportional to the Planck length: it is a remnant of the polymer quantization scheme and is ultimately expected to reflect the underlying discrete, granular structure of space-time at the Planck scale predicted by LQG (we recall that the associated effective Hamiltonian constraint $H_{\text{eff}}^{(3)}$ is related to LQG).

The metric in Eq.~(\ref{eq:ds23}) only approaches the Schwarzschild metric at $r \to \infty$ if $n=0$. In order to ensure phenomenological viability, and in particular agreement with weak-field gravity tests, we will therefore assume $n=0$ in what follows. With this choice, the functions $f(r)$ and $g(r)$ appearing in Eq.~(\ref{eq:ds2}) are given by the following:
\begin{align}
&f(r) = 1-\frac{r^2}{\xi^2}\arcsin \left ( \frac{2GM\xi^2}{r^3} \right )\,,
\label{eq:fr} \\
&g(r) = \left ( 1-\frac{4G^2\xi^4M^2}{r^6} \right ) \left [ 1-\frac{r^2}{\xi^2}\arcsin \left ( \frac{2GM\xi^2}{r^3} \right ) \right ] \,.
\label{eq:gr}
\end{align}
We will refer to the quantum-corrected space-time described by Eqs.~(\ref{eq:ds2},\ref{eq:fr},\ref{eq:gr}) as the ``ZLMY black hole'' (or more generally ZLMY space-time), from the initials of the authors of Ref.~\cite{Zhang:2024ney}. We note that, since $f(r) \neq g(r)$ from Eqs.~(\ref{eq:fr},\ref{eq:gr}), the space-time in question is manifestly non-\textit{tr}-symmetric, falling within the category we studied earlier in Ref.~\cite{Calza:2024xdh}: it is worth noting that space-times of this type appear ubiquitously in quantum gravity approaches to singularity resolution~\cite{Modesto:2008im,Peltola:2008pa,Peltola:2009jm,Bianchi:2018mml,DAmbrosio:2018wgv}.

We now restrict the quantum parameter to the range $\xi/M<\pi\sqrt{\pi}M/\sqrt{2} \approx 3.94\,M$. With this choice, the space-time describes an asymptotically flat BH with horizon located at the radial coordinate $r_H$, determined by solving the following equation:
\begin{equation}
g(r_H)=0\,.
\label{eq:rh}
\end{equation}
The classical singularity is replaced by a traversable wormhole with throat located at $r_{\min}=(2M\xi^2)^{1/3}$, beyond which the metric extends into a Schwarzschild-de Sitter space-time with negative mass~\cite{Zhang:2024ney}. It is worth noting that no Cauchy horizon is present throughout the space-time, which is therefore safe against the mass inflation problem and more generally against perturbative instabilities. To confirm that the space-time is regular, we explicitly calculate three curvature invariants: the Ricci scalar $R \equiv g^{\mu\nu}R_{\mu\nu}$, Ricci tensor squared $R_{\mu\nu}R^{\mu\nu}$, and Kretschmann scalar ${\cal K} \equiv R_{\mu\nu\rho\sigma}R^{\mu\nu\rho\sigma}$. Explicit expressions for these invariants are reported (to the best of our knowledge for the first time) in Appendix~\ref{sec:invariants} and have been verified to be finite for $r>r_{\min}$, while recovering their Schwarzschild limits for $\xi \to 0$: in what follows, we will therefore refer to $\xi$ as the ``regularizing parameter''. In short, the ZLMY space-time effectively describes a regular BH which resolves the classical singularity, is free of Cauchy horizons, and emerges from an effective canonical LQG-related theory which is guaranteed to be generally covariant. This solid theoretical foundation places the ZLMY space-time, and therefore our study, on much firmer footing compared to earlier more phenomenological studies, including our earlier works on the subject~\cite{Calza:2024fzo,Calza:2024xdh}.

\begin{figure}[!t]
\centering
\includegraphics[width=1.0\columnwidth]{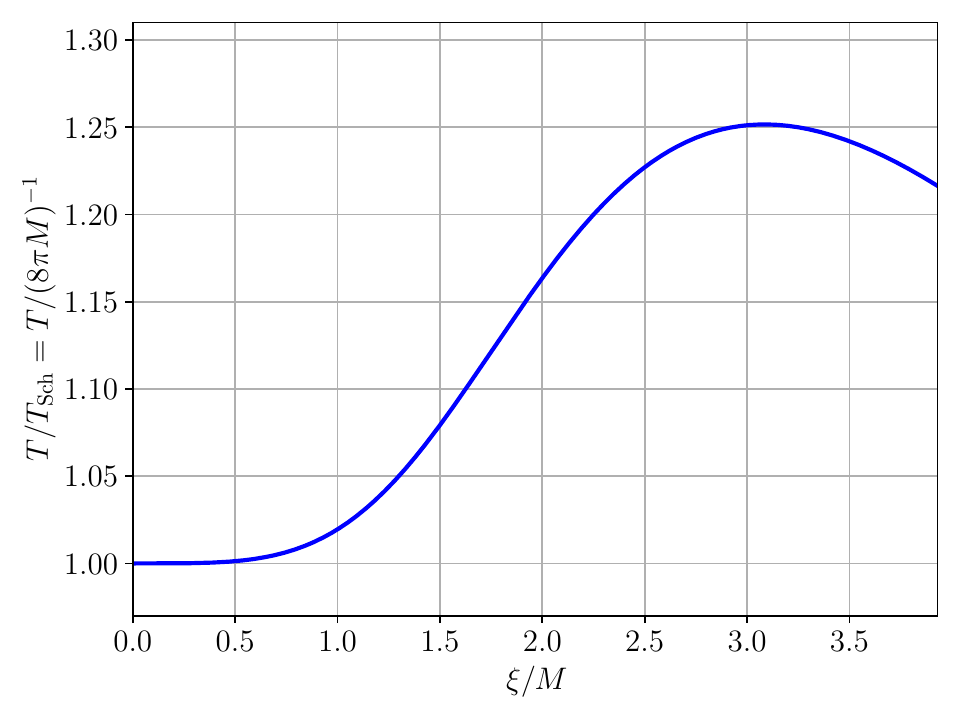}
\caption{Evolution of the Zhang-Lewandowski-Ma-Yang BH temperature as a function of the regularizing parameter $\xi$, in units of mass $M$. The temperature is normalized by that of a Schwarzschild BH of the same mass, $T_{\text{Sch}}=1/8\pi M$. While the temperature is not a monotonically increasing function of $\xi$, it is always larger than the temperature of a Schwarzschild BH of the same mass.}
\label{fig:temperature}
\end{figure}

As discussed earlier, our main phenomenological interest in the ZLMY BH is the fact that its temperature is larger than that of a Schwarzschild BH of the same mass, as long as $\xi \neq 0$. To confirm this, we compute the ZLMY BH temperature, which is given by the following:
\begin{equation}
T=\sqrt{\frac{g(r)}{f(r)}}\frac{f'(r)}{4\pi}\Bigg\vert_{r_H}\,,
\label{eq:temperature}
\end{equation}
with the prime denoting a derivative with respect to $r$. We show the evolution of the temperature as a function of the regularizing parameter $\xi$ in Fig.~\ref{fig:temperature}. From the Figure we see that, although the temperature is not a monotonically increasing function of $\xi$, it is nevertheless always larger than the temperature of a Schwarzschild BH of the same mass, $T_{\text{Sch}}=1/8\pi M$. In particular, the ratio to the Schwarzschild temperature is maximized for $\xi/M \sim 3$, in which case the ZLMY BH is $\approx 25\%$ hotter compared to the Schwarzschild BH.

\section{Methodology}
\label{sec:methodology}

We now briefly discuss the methodology adopted to derive constraints on primordial regular Zhang-Lewandowski-Ma-Yang BHs as DM candidates. The discussion is kept deliberately brief since the methodology is nearly identical to that adopted in the two earlier papers in this series~\cite{Calza:2024fzo,Calza:2024xdh}, except for two small numerical details we will comment on, and we therefore encourage the reader to consult Refs.~\cite{Calza:2024fzo,Calza:2024xdh} for a more detailed description of the methodology. In short, the analysis consists of three steps:
\begin{enumerate}
\item calculating the graybody factors (GBFs) of the ZLMY BH;
\item using the above, as well as the temperature calculated earlier and shown in Fig.~\ref{fig:temperature}, to derive the ZLMY BH evaporation spectra;
\item use these to derive constraints on primordial ZLMY BHs from observations of the diffuse extragalactic $\gamma$-ray background (EGRB).
\end{enumerate}
The final output of the above steps consists of (evaporation) constraints on $f_{\text{pbh}}$, the fraction of DM in the form of ZLMY PBHs, versus the ZLMY PBH mass $M_{\text{pbh}}$. We compute these limits in the $10^{15} \lesssim M_{\text{pbh}}/{\text{g}} \lesssim 10^{18}$ range, and for different values of the regularizing parameter $\xi$, which for definiteness we fix to $\xi/M=1,2,3$ (and, for comparison, also the $\xi/M=0$ case which corresponds to Schwarzschild PBHs).

The GBFs of the ZLMY BH are computed by solving the Teukolsky equation~\cite{Teukolsky:1973ha}, with boundary conditions corresponding to a scattering problem with wave packets originating from the horizon. The energy-dependent transmission coefficient then corresponds to the GBF for the specific spin and mode of the field considered. More specifically, in our earlier papers the radial Teukolsky equation is solved using a method which Ref.~\cite{Arbey:2025dnc} refers to as ``direct method''. The latter makes use of the Frobenius method to iteratively derive the coefficients for the Taylor expansion of the near-horizon solution. This solution is then used as starting point to integrate the equation up to large distances, where the asymptotic behaviour of the solution is well known, and where the GBF can be read off relatively straightforwardly. Using this method, we calculate the GBFs $\Gamma_l^s$ for spin $s=1$ (as we are interested in photons) and $l=1,2,3,4$, verifying that the choice of truncating the calculation at the field angular node number $l=4$ is sufficient for our purposes. We encourage the reader to consult Sec.~IIIA and Appendix~A of Ref.~\cite{Calza:2024fzo}, where the methodology for computing the GBFs is discussed in much more detail.

Methodologically, the present analysis differs from the earlier companion papers~\cite{Calza:2024fzo,Calza:2024xdh} in two minor aspects. Firstly, the calculation of the GBFs is performed using a modified version of the \texttt{GrayHawk} code, developed by one of us and publicly released after our two earlier papers had been published~\cite{Calza:2025whq} (see also Refs.~\cite{Arbey:2025dnc,Yuan:2025eyi}). \texttt{GrayHawk} essentially implements a method which Ref.~\cite{Arbey:2025dnc} refers to as ``Chandrasekhar method''. This method is faster and simpler than the direct method, employed in Refs.~\cite{Calza:2024fzo,Calza:2024xdh}, and, in the case of spherical symmetry, guarantees the same degree of accuracy. In fact, the output of \texttt{GrayHawk} have been explicitly tested against a few benchmark spectra from our earlier papers~\cite{Calza:2024fzo,Calza:2024xdh}, finding excellent agreement.

The second methodological difference concerns an additional numerical step required to solve the Teukolsky equation. It is worth recalling that the Chandrasekhar method relies on the tortoise coordinate $r^{\star}$, defined starting from the following:
\begin{equation}
\frac{dr^{\star}}{dr}=\frac{1}{\sqrt{f(r)g(r)}}\,,
\label{eq:rstar}
\end{equation}
with the metric functions $f(r)$ and $g(r)$ defined in Eq.~(\ref{eq:ds2}). However, to explicitly solve the resulting Schr\"{o}dinger-like radial Teukolsky equation, described by a geometrical potential $V$, explicit knowledge of the \textit{inverse} map $r(r^{\star})$ is required, see e.g.\ Eq.~(12) in Ref.~\cite{Calza:2025whq}. For ZLMY BHs, this inversion needs to be performed numerically since there the following indefinite integral admits no analytical solution:
\begin{equation}
\int\frac{dr}{\sqrt{f(r)g(r)}}={\tilde r}^{\star}(r)+c_1\,.
\label{eq:inversion}
\end{equation}
If such a solution existed, setting $c_1=0$ would allow us to identify the required map $r^{\star}(r)={\tilde r}^{\star}(r)$. To address this problem, we set up a dense grid of radial values $\{r_i\}$ from the event horizon $r=r_H$ to sufficiently large distances. We then numerically perform the following integral between consecutive points on the grid:
\begin{equation}
Y_i \equiv \int^{r_{i+1}}_{r_i}\frac{dr}{\sqrt{f(r)g(r)}}\,.
\end{equation}
This allows us to obtain a numerically reconstructed version of the function $r^{\star}(r)$ as follows:
\begin{equation}
\left\{ \frac{r_{i+1}-r_i}{2}, Y_0 + \sum_{j=1}^i Y_j\right\}_i\,.
\label{eq:series}
\end{equation}
It is important to notice that $1/{\sqrt{f(r)g(r)}}$ is a monotonically decreasing function for $r>r_H$. It follows that $ \left\{\sum_{j=1}^i Y_{j}\right\}_i$ is a positive series starting from values close to $0$ and increasing with labeling index $i$ as $r$ increases away from $r_H$. In Eq.~(\ref{eq:series}), the role of the term $Y_0$ is to ensure that the series correctly represents the functional form of $r^{\star}(r)$, by guaranteeing that the boundary conditions of the integration problem are being adequately set. Specifically, we define $F_{r_H}^{(1)}(r)$ and $G_{r_H}^{(1)}(r)$ as the first-order Taylor expansions of the functions $f(r)$ and $g(r)$ around $r_H$. We then consider the following integral:
\begin{equation}
\int \frac{dr}{\sqrt{F_{r_H}^{(1)}(r)G_{r_H}^{(1)}(r)}}=\tilde R^{\star}(r)+c_2\,.
\end{equation}
When setting $c_2=0$, this integral is a very good approximation $r^{\star}(r)$ sufficiently close to $r_H$. We therefore set $Y_0=\tilde R^{\star}(r_1)$. Finally, the numerical inversion is performed by inverting the orders of the entries in the series given by Eq.~(\ref{eq:series}). We explicitly test this inversion method against known analytic maps $r(r^{\star})$, finding excellent agreement. The method will be implemented in a future version of the \texttt{GrayHawk} code.

\begin{figure}[!t]
\centering
\includegraphics[width=1.0\columnwidth]{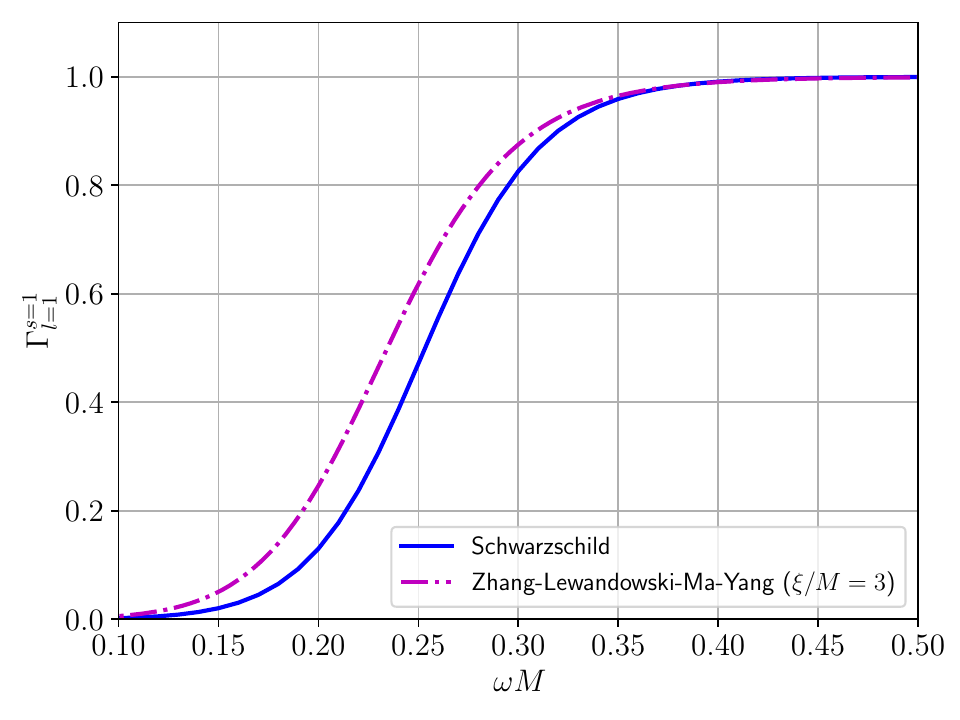}
\caption{Graybody factor $\Gamma_{l=1}^{s=1}$ as a function of $\omega M$ for Schwarzschild BHs (blue solid curve) and Zhang-Lewandowski-Ma-Yang regular BHs (magenta dash-dotted curve). For illustrative purposes, we only plot $\Gamma_{l=1}^{s=1}$, since we are interested in photons ($s=1$) and the dominant emission mode is the $l=1$ one, while fixing the regularizing parameter to $\xi/M=3$. We see that the Zhang-Lewandowski-Ma-Yang GBF is consistently higher than the Schwarzschild one, and starts rising at much lower energies. The features shown in this plot do not change sensibly for higher values of $l$ and other values of $\xi$.}
\label{fig:gbf}
\end{figure}

As an example, in Fig.~\ref{fig:gbf} we plot the $\Gamma_{l=1}^{s=1}$ GBF for the ZLMY BH, with the regularizing parameter set to $\xi/M=3$. We see that the GBF is higher than its Schwarzschild counterpart. In particular, it starts rising at lower energies ($\omega M \lesssim 0.15$), while both plateau at $\omega M \sim 0.4$. Throughout this range, the ZLMY BH GBFs are consistently larger than the Schwarzschild ones. For instance, at $\omega M=0.25$, the ZLMY GBF is $\approx 33\%$ larger than the Schwarzschild one. We have explicitly verified that this trend is consistently recovered for other values of $l$ and $\xi$.~\footnote{The previous two papers in this series~\cite{Calza:2024fzo,Calza:2024xdh} contain a typo when plotting and discussing the GBFs. The quantity on the $x$ axis should have been $\omega r_H$ rather than $\omega/M$. We have checked that this was merely a typo which does not affect our previous results. In the present work, to conform to standard practice in the field, we have chosen to adopt $M$ instead than $r_H$ as normalizing variable, so the GBF plotted in Fig.~\ref{fig:gbf} is given as a function of $\omega M$.}

\begin{figure}[!t]
\centering
\includegraphics[width=1.0\columnwidth]{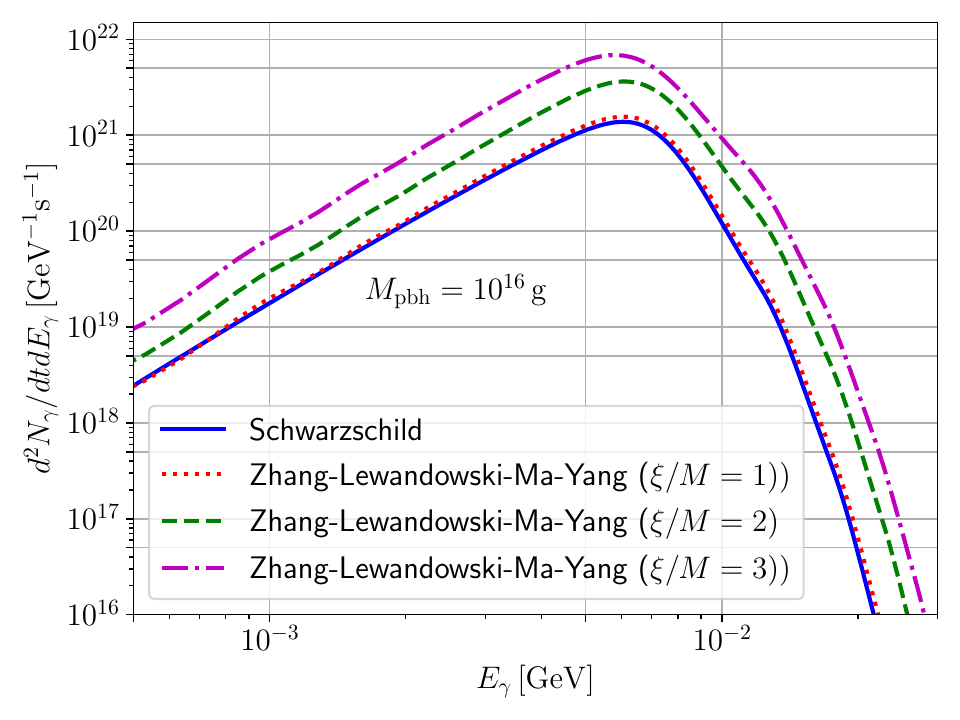}
\caption{Primary photon spectra resulting from the evaporation of a primordial regular Zhang-Lewandowski-Ma-Yang BH of mass $10^{16}\,{\text{g}}$ for different values of the regularizing parameter $\xi$ (normalized by the mass $M$): $\xi/M=1$ (red dotted curve), $2$ (green dashed curve), and $3$ (magenta dash-dotted curve). The blue solid curve corresponds to the case $\xi/M=0$, which recovers the Schwarzschild BH.}
\label{fig:spectraphotonszlmy}
\end{figure}

The GBFs we compute above are used to calculate the primary photon spectra resulting from Hawking evaporation of ZLMY PBHs. Under the reasonable assumption that photons are not coupled to the regularizing parameter $\xi$, the number of photons with energy $E_{\gamma}$ emitted per unit time per unit energy is given by the following:
\begin{equation}
\frac{d^2N_{\gamma}}{dtdE_{\gamma}}(E_{\gamma},\xi)=\frac{1}{\pi}\sum_{l,m}\frac{\Gamma^s_{l,m}(E_{\gamma},\xi)}{e^{E_{\gamma}/T(\xi)}-1}\,.
\label{eq:d2ndtdei}
\end{equation}
The resulting photon spectra are shown in Fig.~\ref{fig:spectraphotonszlmy} for a representative ZLMY PBH of mass $M_{\text{pbh}}=10^{16}\,{\text{g}}$, a value which is located roughly halfway within the mass range of interest to us, and for the same values of the regularizing parameter $\xi/M=1,2,3$ discussed earlier. We see that in all three cases the emission spectra are stronger than their Schwarzschild counterparts, especially at low and intermediate energies ($E_{\gamma} \lesssim 10\,{\text{MeV}}$), and peak around $7\,{\text{MeV}}$. Although we have chosen $M_{\text{pbh}}=10^{16}\,{\text{g}}$ for representative purposes, these generic characteristics do not depend on the value of $M_{\text{pbh}}$.

The fact that the emission from Hawking evaporation of PBHs of a given mass is stronger for ZLMY BHs compared to their Schwarzschild counterparts is in line with expectations. In fact, we have seen that both the temperature (Fig.~\ref{fig:temperature}) and GBFs (Fig.~\ref{fig:gbf}) of these BHs increase relative to the Schwarzschild case. Inspecting Eq.~(\ref{eq:d2ndtdei}), we can see how these increases both go in the direction of enhancing the resulting spectra, although the dominant effect is expected to be that of the increase in temperature, given the exponential (linear) dependence of $d^2N_{\gamma}/dtdE_{\gamma}$ on the temperature (GBFs).

From the evaporation spectrum of a primordial ZLMY BH of mass $M_{\text{pbh}}$, we then compute the rate (per unit time per unit area per unit solid angle) of emitted photons whose present-day energy is $E_{\gamma 0}$ as follows:
\begin{align}
I(E_{\gamma 0})= &\frac{c}{4\pi}n_{\text{pbh}}(t_0)E_{\gamma 0} \nonumber \\
\times&\int^{z_{\star}}_{0} \frac{dz}{H(z)}\frac{d^2{N}_{\gamma}}{dt dE_{\gamma}}(M_{\text{pbh}},(1+z)E_{\gamma 0})\,,
\label{eq:flux}
\end{align}
where the integral extends to the recombination redshift $z_{\star}$, and $H(z)$ denotes the expansion rate. For consistency, we use the same flat $\Lambda$CDM cosmological model adopted in our previous works~\cite{Calza:2024fzo,Calza:2024xdh}. In Eq.~(\ref{eq:flux}), the only unknown parameter is then the current comoving number density of primordial ZLMY BHs, $n_{\text{pbh}}(t_0)$. For consistency with our two companion papers~\cite{Calza:2024fzo,Calza:2024xdh}, we set upper limits on this quantity using measurements of the EGRB in the $0.1 \lesssim E_{\gamma}/{\text{MeV}} \lesssim 100$ range from the HEAO-1, COMPTEL, and EGRET telescopes~\cite{Gruber:1999yr,Schoenfelder:2000bu,Strong:2004ry}. We determine the maximum allowed value of $n_{\text{pbh}}(t_0)$ as being the one such that the theoretical prediction in Eq.~(\ref{eq:flux}) first overshoots one of the measured EGRB datapoints. We refer the reader to the discussion in Ref.~\cite{Calza:2024fzo} and Fig.~6 therein for further details on this method which, while simplified, has been adopted in the majority of the works in the PBH evaporation literature, including the seminal Ref.~\cite{Carr:2009jm}. We then translate a limiting value $\widetilde{n}_{\text{pbh}}(t_0)$ into a limiting value $\widetilde{f}_{\text{pbh}}$ as follows:
\begin{equation}
\widetilde{f}_{\text{pbh}}(M_{\text{pbh}}) \equiv \frac{\Omega_{\text{pbh}}}{\Omega_{\text{dm}}} = \frac{\widetilde{n}_{\text{pbh}}(t_0)M_{\text{pbh}}}{\rho_{\text{crit},0}\Omega_{\text{dm}}}\,,
\label{eq:npbhtofpbh}
\end{equation}
where $\rho_{\text{crit},0}=3H_0^2/8\pi G$ is the present-day critical density in terms of the Hubble constant $H_0$, which we set to $H_0=72\,{\text{km}}/{\text{s}}/{\text{Mpc}}$, while $\Omega_{\text{dm}}$ is the density parameter of DM, set to $\Omega_{\text{dm}}=0.212$. We stress that our results are essentially insensitive to the choice of cosmological parameters as long as these are sufficiently close to the best-fit $\Lambda$CDM ones, even when taking into account significant uncertainties in certain cosmological parameters (e.g.\ $H_0$) due to cosmological tensions~\cite{CosmoVerseNetwork:2025alb}.

We note that we are implicitly adopting the same approximations as in our earlier works~\cite{Calza:2024fzo,Calza:2024xdh}, namely:
\begin{enumerate}
\item assuming that primordial ZLMY BHs are isotropically distributed on large scales and cluster in the galactic halo in the same way as other forms of DM;
\item computing only the primary spectrum;
\item assuming a monochromatic mass distribution.
\end{enumerate}
We refer the reader to the discussion in Sec.~IIIC of Ref.~\cite{Calza:2024fzo} for further details on the validity of these approximations and/or limitations thereof, and why these approximations (as well as the simple procedure for setting upper limits on $\widetilde{f}_{\text{pbh}}$) are justified given the scope of the present work. In particular, we are mostly interested in the relative shift in constraints on $f_{\text{pbh}}$ compared to the Schwarzschild case, a quantity which we expect to be only weakly affected by our approximations. Additionally, these approximations also allow for a more direct comparison not only to other works in the literature, but importantly to our companion papers~\cite{Calza:2024fzo,Calza:2024xdh}. We note that the impact of going beyond the monochromatic mass distribution, commonly adopted in the literature, has been extensively discussed in other works~\cite{Kuhnel:2015vtw,Kuhnel:2017pwq,Carr:2017jsz,Raidal:2017mfl,Bellomo:2017zsr,Lehmann:2018ejc,Carr:2018poi,Gow:2019pok,DeLuca:2020ioi,Gow:2020cou,Ashoorioon:2020hln,Bagui:2021dqi,Mukhopadhyay:2022jqc,Papanikolaou:2022chm,Cai:2023ptf}.

\section{Results and discussion}
\label{sec:results}

\begin{figure}[!t]
\centering
\includegraphics[width=1.0\columnwidth]{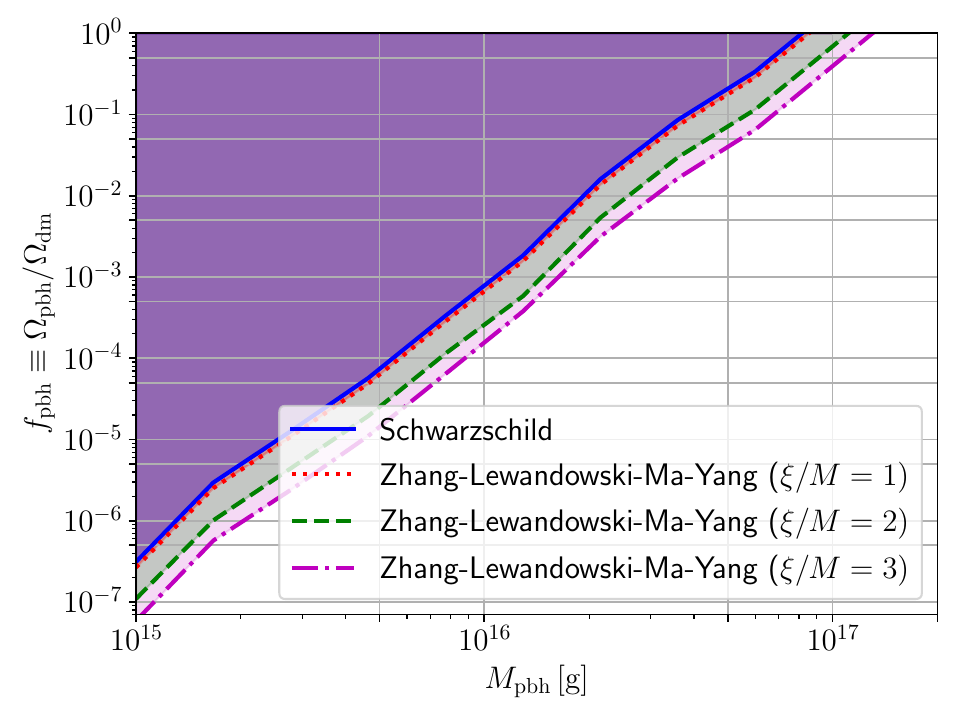}
\caption{Upper limits on $f_{\text{pbh}}$, the fraction of dark matter in the form of primordial regular Zhang-Lewandowski-Ma-Yang BHs, as a function of the PBH mass $M_{\text{pbh}}$. The limits are derived for different values of the regularizing parameter $\xi$ (normalized by the mass $M$), with the shaded regions excluded: $\xi/M=1$ (red dotted curve), $2$ (green dashed curve), and $3$ (magenta dash-dotted curve). Note that the blue solid curve corresponds to the case $\xi/M=0$, which recovers the Schwarzschild BH, whereas the value of $M_{\text{pbh}}$ corresponding to the upper right edge of the $f_{\text{pbh}}$ constraints marks the lower edge of the asteroid mass window.}
\label{fig:fpbhlimitszlmy}
\end{figure}

The limits we derive on the fraction of DM in the form of ZLMY PBHs $f_{\text{pbh}}$, as a function of their mass $M_{\text{pbh}}$, are reported in Fig.~\ref{fig:fpbhlimitszlmy}. In particular, we report the constraints obtained considering three different values of the regularizing parameter $\xi$: $\xi/M=1$ (red dotted curve), $2$ (green dashed curve), and $3$ (magenta dash-dotted curve). The black curve instead corresponds to the limit $\xi/M \to 0$, where the metric in Eq.~(\ref{eq:ds23}) recovers that of the Schwarzschild BH: we explicitly checked that our constraints in this limit recover those of Ref.~\cite{Carr:2009jm}. It is also worth noting that the lower edge of the asteroid mass window, for a given value of $\xi$, is set by the value of $M_{\text{pbh}}$ where the limit $f_{\text{pbh}}<1$ is saturated. In the Schwarzschild PBH case, and within the approximations adopted, this lower edge is given by $\simeq 7 \times 10^{16}\,{\text{g}}$.

We find that, at a given $M_{\text{pbh}}$, increasing $\xi$ leads to tighter constraints on $f_{\text{pbh}}$, precisely as one could have expected, given the stronger emission spectra observed in Fig.~\ref{fig:spectraphotonszlmy}, themselves a reflection of the higher temperature (see Fig.~\ref{fig:temperature}) and larger GBFs (see Fig.~\ref{fig:gbf}) discussed earlier. For $\xi/M=3$, which roughly maximizes the temperature of the ZLMY BH (see the peak in Fig.~\ref{fig:temperature}, where the temperature is roughly $25\%$ larger than the Schwarzschild one), we find that the constraints on $f_{\text{pbh}}$ at fixed $M_{\text{pbh}}$ are approximately an order of magnitude tighter compared to those for Schwarzschild PBHs. Although we have not shown this in the Figures, further increasing $\xi$ would actually make the constraints weaken relative to the $\xi/M=3$ case, because of the non-monotonic behaviour of the temperature, which decreases for $\xi/M \gtrsim 3$. At any rate, these shifts result in the lower edge of the asteroid mass window moving towards \textit{larger} masses. For $\xi/M=3$, we find that this quantity increases to $\simeq 2 \times 10^{17}\,{\text{g}}$, thereby moving by about half an order of magnitude compared to the Schwarzschild case. The range of parameter space where primordial regular ZLMY BHs can make up all the DM, consistent with the non-observation of the products of their Hawking evaporation, is therefore reduced compared to the parameter space for Schwarzschild BHs.

Five comments are in order at this point. Firstly, we note that the results reported go completely in the opposite direction compared to our two earlier works~\cite{Calza:2024fzo,Calza:2024xdh}, where we observed weaker constraints on $f_{\text{pbh}}$ and thereby a broadening of the asteroid mass window. Although this was probably not sufficiently emphasized in Refs.~\cite{Calza:2024fzo,Calza:2024xdh}, the evaporation constraints for the metrics considered there can in fact be completely \textit{eliminated} if the regularizing parameter reaches its extremal value, in which case the regular BH temperature reaches zero, therefore leading to a non-evaporating relic (see also Ref.~\cite{Davies:2024ysj}). It is nevertheless worth reminding the reader that the metrics considered earlier were all phenomenological or, at best, quantum gravity-inspired. The metric we consider here is instead an \textit{exact} solution of a generally covariant effective canonical (loop) quantum gravity model, which is not only non-singular but also free of Cauchy horizons, making it a serious candidate to address the singularity problem in a theoretically healthy way. The constraints on $f_{\text{pbh}}$ derived here are therefore arguably somewhat more robust and theoretically well-grounded compared to the earlier ones.

Next, as in our earlier works~\cite{Calza:2024fzo,Calza:2024xdh}, we are considering spherically symmetric space-times, i.e. non-rotating BHs. In the Schwarzschild case, considering the PBH spin and therefore the Kerr metric has been demonstrated to lead to tighter constraints on $f_{\text{pbh}}$, thereby leading to a smaller asteroid mass window~\cite{Arbey:2019vqx,Arbey:2020yzj}. The main reason is that, even though increasing the Kerr BH spin lowers its temperature, the GBF associated to the maximally co-rotating $m$ mode starts increasing at much smaller energies compared to the other modes, leading to an overall enhancement of the Hawking evaporation spectrum. By thermodynamical considerations, it is more than plausible that including rotation in the ZLMY BH would lead to a similar effect, and therefore that the asteroid mass window would be closed down even further compared to what is shown in Fig.~\ref{fig:fpbhlimitszlmy}. However, checking this explicitly would require one to construct the rotating version of Eq.~(\ref{eq:ds23}), a task to which the Newman-Janis algorithm is not well-suited given the non-phenomenological nature of the metric. We therefore defer this interesting question to follow-up work.~\footnote{In a similar spirit, it might be interesting to assess whether potential new windows for small $M_{\text{pbh}} \lesssim 10^{15}\,{\text{g}}$ PBHs as DM candidates opened up by the memory burden effect, which suppresses the evaporation of BHs once about half of their mass has been lost~\cite{Dvali:2020wft}, may be altered (relative to the Schwarzschild case) when working with primordial regular BHs.}

Moreover, in deriving our constraints on $f_{\text{pbh}}$ versus $M_{\text{pbh}}$, we are implicitly making a quasi-static approximation, denoting by $M_{\text{pbh}}$ the mass of primordial ZLMY BHs both at formation and today. In other words, we are assuming that these BHs have only lost a negligible fraction of their mass throughout the age of the Universe. This assumption is valid for Schwarzschild PBHs if $M_{\text{pbh}} \gtrsim 10^{15}\,{\text{g}}$, i.e.\ the mass range of interest to us: in this case, it can be shown that these BHs have only lost a sub-\% fraction of their mass from formation up to the present time, as their lifetime is much larger than the age of the Universe (see e.g.\ Appendix~B of Ref.~\cite{Calza:2024fzo}). However, whether this approximation is still a good one needs to be explicitly demonstrated for primordial ZLMY BHs, as they are hotter and emit more strongly than their Schwarzschild counterparts (see Fig.~\ref{fig:temperature} and Fig.~\ref{fig:spectraphotonszlmy}), and are therefore expected to have a shorter lifetime compared to these.

To check this, we perform the same calculation detailed in Appendix~B of the first paper in this series~\cite{Calza:2024fzo}, to which we refer the reader for more details. Specifically, we first calculate the depletion function for fields of arbitrary spin, and evaluate the BH mass loss rate~\cite{Page:1976df,Page:1976ki,Page:1977um}. Backward-integrating the mass loss equation setting the boundary condition $M(t_U)=0$, with $t_U$ being the age of the Universe, allows us to determine $M(t=0)$, i.e.\ the mass of a BH whose lifetime is the age of the Universe. For Schwarzschild PBHs, we find $M(t=0) \sim 5 \times 10^{14}\,{\text{g}}$, in excellent agreement with earlier works~\cite{Page:1976df}. For ZLMY BHs, we consider the case where $\xi/M=3$, which roughly maximizes its temperature, and therefore minimizes its lifetime. Even in this most conservative case, we find the mass of a ZLMY BH whose lifetime is the age of the Universe to be $M(t=0) \sim 7.5 \times 10^{14}\,{\text{g}}$, only slightly larger than its Schwarzschild counterpart.

Our focus is on the $M_{\text{pbh}} \gtrsim 10^{15}\,{\text{g}}$ mass range, so a relevant question is what fraction of its initial mass has a primordial ZLMY BH within this mass range lost throughout the age of the Universe. For $M(t=0)=10^{15}\,{\text{g}}$, we find that a primordial ZLMY BH with $\xi/M=3$ has a lifetime of $\sim 3\,t_U$, and has lost $\lesssim 10\%$ of its mass by $t \sim t_U$. Since in this work we are mostly interested in the lower edge of the asteroid mass range, i.e.\ the mass $M_{\text{pbh}}$ at which the overclosure constraint $f_{\text{pbh}}<1$ is saturated, we now consider a primordial ZLMY BH with $\xi/M=3$ and $M(t=0)=10^{17}\,{\text{g}}$. In this case, we find a lifetime of $\sim 3 \times 10^6\,t_U$, and that such a PBH has lost $\ll 1\%$ of its mass by $t \sim t_U$. We conclude that, within the mass range of interest, primordial ZLMY BHs are far from having fully evaporated and have only lost a small fraction of their initial mass, thereby justifying our quasi-static approximation.

Additionally, we remind the reader of a number of caveats surrounding our results, already discussed in Refs.~\cite{Calza:2024fzo,Calza:2024xdh}. Firstly, we are treating the parameter $\xi/M$ as an ``universal hair'' (in the language of Ref.~\cite{Vagnozzi:2022moj}), but whether or not this is motivated from a fundamental perspective is an open question which should eventually be addressed. In addition, our statements on the asteroid mass window shrinking are contingent on the upper edge thereof, which is determined by lensing constraints, remaining fixed. As argued earlier~\cite{Calza:2024fzo,Calza:2024xdh}, we do indeed expect this to be the case, although a detailed study of microlensing constraints on primordial regular BHs (including primordial ZLMY BHs) will be the subject of a separate work. Finally, non-evaporation constraints on $f_{\text{pbh}}$ (e.g.\ dynamical, accretion, and CMB limits, see Ref.~\cite{Carr:2023tpt} for a recent summary) are expected to be relevant within a completely different mass range, and will be the subject of a follow-up work.

Beyond these phenomenological aspects, it is worth briefly commenting on the broader and more general cosmological and astrophysical implications of PBH DM. On large scales, a population of PBHs (singular or regular) behaves dynamically as a pressureless, collisionless medium. The discrete nature of the PBH distribution, while potentially leaving interesting imprints, becomes irrelevant once averaged over many objects: therefore, prior to horizon re-entry, PBHs generate gravitational potentials which set the stage for the later growth of baryon perturbations, in complete analogy to standard cold DM. In this respect, we note that the cosmological evolution driven by primordial regular BHs should be practically indistinguishable from that produced by their singular counterparts.

The discrete nature of the PBH distribution may nevertheless lead to interesting astrophysical signatures. Such a discreteness is inevitably accompanied by Poisson fluctuations, which introduce additional (shot noise-like) small-scale power. This would influence the angular momentum evolution of early halos as well as their collapse dynamics and could, at least qualitatively, be related to the long-standing galactic spin problem. In addition, the clustering of PBHs on galactic scales could plausibly affect the observed morphology-density relation of galaxies. For instance, denser PBH environments might favor merger-driven, pressure-supported systems such as elliptical and lenticular galaxies, with the converse true in less dense environments, which may aid the survival of rotationally supported disks, thereby leading to a preferential formation of (flocculent) spiral and irregular galaxies. While these deliberately qualitative cosmological and astrophysical considerations lie beyond the scope of this work, they show how the phenomenology of primordial (singular or regular) BHs, including ZLMY PBHs, fit into the broader cosmological and astrophysical picture.

\section{Conclusions}
\label{sec:conclusions}

A central challenge in theoretical physics remains that of finding a consistent framework of quantum gravity, with an important requirement for theoretical and phenomenological consistency being that of maintaining general covariance. Could such a consistent framework simultaneously offer solutions to two other major challenges in physics, such as the dark matter and singularity problems? This is the question we seek to address in the present study which, building on our earlier results~\cite{Calza:2024fzo,Calza:2024xdh}, marks the third entry in our ongoing line of work on primordial regular black holes as DM candidates: such a line of research is motivated by the fact that almost all studies treat PBHs as Schwarzschild or Kerr BHs, both of which feature pathological singularities. To tie together the aforementioned themes, we turn our attention to the novel (non-singular) BH solution presented in Ref.~\cite{Zhang:2024ney}, referred to as the Zhang-Lewandowski-Ma-Yang (ZLMY) BH, which stands out for being free of Cauchy horizons, and for emerging from a framework which explicitly enforces general covariance within LQG-related effective canonical QG theories: these key features place this work on a theoretically more robust footing compared to our earlier studies~\cite{Calza:2024fzo,Calza:2024xdh}. In addition, unlike in these earlier works, the ZLMY BH temperature increases relative to the Schwarzschild case, leading to distinct phenomenological predictions.

We find that this increase in temperature (at a given BH mass) is responsible for stronger Hawking evaporation spectra, in turn leading to tighter constraints on $f_{\text{pbh}}$, the fraction of DM in the form of primordial regular ZLMY BHs. This directly results in a smaller asteroid mass window where ZLMY PBHs could make up the entire DM of the Universe: in particular, the lower edge of the asteroid mass window moves up by approximately half an order of magnitude in the most extreme case. This behaviour is exactly the opposite of what we observed in our earlier works~\cite{Calza:2024fzo,Calza:2024xdh}, where evaporation constraints could in principle be completely bypassed as the regularizing parameter approaches its extremal limit. However, we recall that our earlier works were mostly based on phenomenological space-times. It should be noted that working within a consistent theoretical framework ensuring general covariance, regularity, absence of Cauchy horizons, and connection to a promising QG framework actually reduces the parameter space where PBHs can constitute the DM.

Together with our earlier works, our results further support the idea that there is a promising common ground between the DM, singularity, and QG problems, worthy of further exploration. As remarked in Refs.~\cite{Calza:2024fzo,Calza:2024xdh}, this series of works should be intended as pilot studies in this exciting direction, as several open problems remain. For instance, an important open question is whether the formation mechanism of primordial regular BHs is significantly different from the Schwarzschild case, as departures from the assumptions underlying Birkhoff's theorem may entail a non-unique endpoint for gravitational collapse: this may potentially lead to complementary observational signatures, which may offer new routes towards testing the regular nature of astrophysical BHs. Another key task is to revisit non-evaporation (lensing, accretion, dynamical) constraints on primordial regular BHs: although we have loosely argued that we do not expect these to change significantly, such a statement remains to be checked explicitly. Finally, we note that the same canonical polymerized framework underlying the ZLMY space-time, which bounds curvature invariants and cures future spacelike singularities, might also tame future timelike ones (of more cosmological interest)~\cite{deHaro:2023lbq}: indeed, works on loop quantum cosmology have shown how future finite-time singularities (e.g.\ Big Rip/Type I, as well as often Type III) are generally avoided in these scenarios, whereas ``sudden'' Type II singularities may or may not be avoided, with the answer being quite sensitive to the polymerization scheme~\cite{Bojowald:2001xe,Singh:2006im,Cailleteau:2008wu,Ashtekar:2008ay,Bamba:2012ka,Odintsov:2014gea,Odintsov:2016tar,Chinaglia:2017wim,Odintsov:2018awm,Bojowald:2025ocr}. It could also be worth extending these studies to modified gravity models such as $f(R)$ gravity, although the implications for cosmological singularities are likely to depend strongly on the associated effective Hamiltonian and polymerization scheme. While we defer all these points to follow-up work, we stress once more a general but important message of the present work: even ideas or frameworks which at first may appear remote or speculative -- such as LQG -- can have important consequences for both astrophysical observations and other open problems in physics, whereas the issue of a theoretically consistent formulation can make a significant difference at the level of phenomenological predictions.

\begin{acknowledgments}
\noindent We are grateful to Sergio Zerbini for many helpful discussions and for alerting us to the ZLMY BH. We acknowledge support from the Istituto Nazionale di Fisica Nucleare (INFN) through the Commissione Scientifica Nazionale 4 (CSN4) Iniziativa Specifica ``Quantum Fields in Gravity, Cosmology and Black Holes'' (FLAG). M.C., G.-W.Y. and S.V. acknowledge support from the University of Trento and the Provincia Autonoma di Trento (PAT, Autonomous Province of Trento) through the UniTrento Internal Call for Research 2023 grant ``Searching for Dark Energy off the beaten track'' (DARKTRACK, grant agreement no.\ E63C22000500003). This publication is based upon work from the COST Action CA21136 ``Addressing observational tensions in cosmology with systematics and fundamental physics'' (CosmoVerse), supported by COST (European Cooperation in Science and Technology).
\end{acknowledgments}

\clearpage

\vspace{1cm} 

\appendix

\section{Curvature invariants for the Zhang-Lewandowski-Ma-Yang black hole}
\label{sec:invariants}

For completeness, here we report explicit expressions for the main curvature invariants of the ZLMY BH, as a function of the regularizing parameter $\xi$. We begin from the Ricci scalar, given by the following expression:
\begin{widetext}
\begin{align}
R \equiv g^{\mu\nu}R_{\mu\nu} = \left ( \dfrac{12}{\xi^2}+\dfrac{24M^2\xi^2}{r^6} \right ) \arcsin \left ( \dfrac{2M\xi^2}{r^3} \right ) -\dfrac{8M \left ( 5M\xi^4+3\sqrt{1-\dfrac{4M^2\xi^4}{r^6}}\, r^5 \right ) }{r^8}\,.
\label{eq:ricci}
\end{align}
\end{widetext}
The Kretschmann scalar is instead given by the following:
\small
\begin{widetext}
\begin{align}
{\cal K} \equiv &R_{\mu\nu\rho\sigma}R^{\mu\nu\rho\sigma} \nonumber \\
=&\quad 24 \left ( \dfrac{1}{\xi^4}+\dfrac{40M^4\xi^4}{r^{12}}+\dfrac{4M^2}{r^6} \right ) \arcsin^2 \left ( \dfrac{2M\xi^2}{r^3} \right ) +\dfrac{16 \left [ 36M^3\xi^4 \sqrt{1-\dfrac{4M^2\xi^4}{r^6}}r^5+9M^2r^{10}+4M^4 \left ( 19\xi^8-9 \xi^4 r^4 \right ) \right ] }{r^{16}} \nonumber \\
-&\quad\dfrac{32M \left ( 52M^3\xi^8+6 M^2\xi^4\sqrt{1-\dfrac{4M^2\xi^4}{r^6}}r^5+5M\xi^4r^6+3\sqrt{1-\dfrac{4M^2\xi^4}{r^6}}r^{11} \right ) }{\xi^2 r^{14}}\arcsin \left ( \dfrac{2M\xi^2}{r^3} \right ) \,.
\label{eq:kretschmann}
\end{align}
\end{widetext}
\normalsize
Finally, the Ricci tensor squared is given by the following:
\small
\begin{widetext}
\begin{align}
R_{\mu\nu}R^{\mu\nu} = & \quad \dfrac{1}{\xi^4 r^{16}} \Bigg \{-48M\xi^2\arcsin \left ( \dfrac{2M\xi^2}{r^3} \right ) r^2 \left ( 24M^3\xi^8+6M^2\xi^4\sqrt{1-\dfrac{4M^2\xi^4}{r^6}}r^5+5M\xi^4r^6+3\sqrt{1-\dfrac{4M^2\xi^4}{r^6}}r^{11} \right ) \nonumber \\
+&\quad 36\arcsin^2 \left ( \dfrac{2M\xi^2}{r^3} \right ) r^4 \left ( 16M^4\xi^8+4M^2\xi^4r^6+r^{12} \right ) + 16M^2\xi^4 \left [ 30M\xi^4\sqrt{1-\dfrac{4M^2\xi^4}{r^6}}r^5+9r^{10}+4M^2 \left ( 11\xi^8-9\xi^4r^4 \right ) \right ] \Bigg \} \,.
\label{eq:riccisquared}
\end{align}
\end{widetext}
\normalsize
We have explicitly checked that the above expressions reduce to the Schwarzschild limits (respectively $R=0$, ${\cal K}=48M^2/r^6$, and $R_{\mu\nu}R^{\mu\nu}=0$) in the $\xi \to 0$ limit where LQG effects vanish.

\clearpage

\bibliography{prbhiii}

\end{document}